\documentclass[10pt]{emulateapj}
\usepackage{apjfonts}
\usepackage{graphicx}
\usepackage{psfig}

\newcommand{\begit}{\begin{itemize}}
\newcommand{\enit}{\end{itemize}}
\newcommand{\begen}{\begin{enumerate}}
\newcommand{\enen}{\end{enumerate}}
\newcommand{\beq}{\begin{equation}}
\newcommand{\eeq}{\end{equation}}

\newcommand{\beqa}{\begin{eqnarray}}
\newcommand{\eeqa}{\end{eqnarray}}

\begin{document}

\title{The Starburst Contribution to the Extra-galactic $\gamma$-Ray
Background}

\author{Todd A.~Thompson,\altaffilmark{1,2}
Eliot Quataert,\altaffilmark{3}
\& Eli Waxman\altaffilmark{4}}

\altaffiltext{1}{
Department of Astrophysical Sciences, Peyton Hall-Ivy Lane,
Princeton University, Princeton, NJ 08544; thomp@astro.princeton.edu}
\altaffiltext{2}{Lyman Spitzer Jr.~Fellow}
\altaffiltext{3}{Astronomy Department
\& Theoretical Astrophysics Center, 601 Campbell Hall,
The University of California, Berkeley, CA 94720;
eliot@astro.berkeley.edu}
\altaffiltext{4}{Physics Faculty, Weizmann Institute of Science, Rehovot
76100, Israel;
waxman@wicc.weizmann.ac.il}

\begin{abstract}

Cosmic ray protons interacting with gas at the mean density of the
interstellar medium in starburst galaxies lose energy rapidly via
inelastic collisions with ambient nuclei.  The resulting pions produce
secondary electrons and positrons, high-energy neutrinos, and
$\gamma$-ray photons.  We estimate the cumulative $\gamma$-ray
emission from starburst galaxies.  We find a total integrated background
above 100 MeV
of $F_\gamma\approx10^{-6}$\,\,GeV\,\,cm$^{-2}$\,\,s$^{-1}$
sr$^{-1}$ and a corresponding specific intensity at GeV
energies of $\nu I_\nu \approx
10^{-7}$\,\,GeV\,\,cm$^{-2}$\,\,s$^{-1}$\,\,sr$^{-1}$.
Starbursts may thus account for a significant fraction of
the extra-galactic $\gamma$-ray background.  We show that the
FIR-radio correlation provides a strong constraint on the $\gamma$-ray
emission from starburst galaxies because pions decay into both
$\gamma$-rays and radio-emitting electron/positron pairs.  We identify
several nearby systems where the potential for observing $\gamma$-ray
emission is the most favorable (M82, NGC 253, \& IC 342), predict their
fluxes,
and predict a linear FIR-$\gamma$-ray correlation for the densest
starbursts.  If established, the FIR-$\gamma$-ray correlation would
provide strong evidence for the ``calorimeter'' theory of the
FIR-radio correlation and would imply that cosmic rays in starburst
galaxies interact with gas at approximately the mean density of the
interstellar medium (ISM), thereby providing an important constraint
on the physics of the ISM in starbursts.

\end{abstract}


\section{Introduction}

The magnitude of the extra-galactic $\gamma$-ray background is
uncertain, primarily due to the presence of foreground contaminants
(compare Keshet et al.~2004, Strong et al.~2004, Sreekumar et al.~1998).
{\it GLAST\,}\footnote{http://glast.gsfc.nasa.gov/;
http://www-glast.stanford.edu/} will have an order of magnitude better
sensitivity at GeV energies than previous experiments and should
provide important constraints on the unresolved extra-galactic
$\gamma$-ray emission.  A number of potential contributors to the GeV
$\gamma$-ray background have been discussed in the literature
including blazars (Stecker \& Salamon 1996), galaxy clusters and
groups (Colafrancesco \& Blasi~1998; Dar \& Shaviv~1995),
intergalactic shocks and structure formation (Loeb \& Waxman 2000;
Keshet et al.~2003; Miniati 2002), dark matter annihilation (e.g., Ullio et al.~2002;
Els{\"a}sser \& Mannheim 2005), and normal star-forming galaxies
(Pavlidou \& Fields 2002).  In this paper we calculate the
contribution of starburst galaxies to the GeV $\gamma$-ray background.
In particular, we assess the possibility that starbursts are cosmic
ray proton calorimeters: the dense ISM of these systems acts as a beam
dump for the total energy injected by supernovae into cosmic ray
protons, with a portion of the proton energy emerging as $\gamma$-rays
and high-energy neutrinos.

In \S\ref{section:calorimeter} we review V\"olk's (1989) electron
calorimeter model for the observed FIR-radio correlation (see Thompson
et al.~2006, hereafter T06).  We then argue that starbursts may also
be proton calorimeters.  In \S\ref{section:correlations}, we predict
the $\gamma$-ray flux from starburst galaxies and highlight which
nearby systems are most likely to have detectable $\gamma$-ray fluxes.
We argue that the observed FIR-radio correlation provides important
constraints on the $\gamma$-ray emission from starbursts because pions
from inelastic proton-proton collisions produce both $\gamma$-rays and
secondary electrons and positrons, which then produce radio synchrotron.
Section
\ref{section:background} then discusses the cumulative starburst
contribution to the extra-galactic $\gamma$-ray background.  Estimates
similar to those presented here for the high-energy neutrino
background from starbursts have recently been made by Loeb \& Waxman
(2006).

Although the $\gamma$-ray flux from individual starbursts has been
estimated by several authors (e.g., Torres 2004, Torres et al.~2004,
Cillis et al.~2005, Paglione et al.~1996, and Blom et al.~1999), this
paper estimates the expected extra-galactic background, defines the
requisite conditions for proton calorimetry, and explicitly connects
the $\gamma$-ray predictions with constraints from the FIR-radio
correlation.

\section{Starburst Galaxies as Cosmic Ray Calorimeters}
\label{section:calorimeter}

Observations of both normal star-forming and starburst galaxies reveal a
linear
correlation between their radio (primarily synchrotron) and FIR
luminosities (van
der Kruit 1971, 1973; Helou et al.~1985; Condon 1992; Yun et al.~2001;
Murgia et al.~2005; Murphy et al.~2006).
V\"olk (1989) argued that the FIR-radio correlation
arises because the synchrotron cooling time for relativistic electrons
generated in supernova shocks is significantly shorter than their
escape time from the galactic disk (the ``calorimeter theory'').  In
this limit, the cosmic ray electrons radiate nearly all of the energy
supplied to them by supernovae. The calorimeter theory is a
particularly compelling explanation for the FIR-radio correlation in
starburst galaxies, where the synchrotron and inverse Compton (IC) cooling
times of cosmic ray electrons are much shorter than in normal
star-forming galaxies (T06).

A priori, one of the strongest objections to the calorimeter
theory is that the radio spectra of galaxies are not compatible with
the expected steepening due to strong synchrotron cooling (e.g.,
Condon 1992; Niklas \& Beck 1997): 
in the presence of strong cooling, models predict $F_\nu
\propto \nu^{-\alpha}$ with $\alpha \approx 1$, while observations
find $\alpha \approx 0.75$, even in starbursts (e.g., Condon et
al.~1991; Condon 1992; Niklas et al.~1997). T06 argue that if
cosmic ray electrons interact with gas at approximately the mean
density of the ISM, ionization and relativistic bremsstrahlung losses
will systematically flatten the nonthermal spectra of starburst
galaxies and simultaneously maintain the linearity of the FIR-radio
correlation.  This strongly supports V\"olk's calorimeter model for
the FIR-radio correlation.  In the model of T06, the conclusion that
cosmic rays interact with gas at roughly the mean density of the ISM
is required to account for the radio spectra of starbursts in the
presence of strong synchrotron cooling.  This requirement has an
important consequence for cosmic ray protons, accelerated together
with electrons in supernova remnants.

If cosmic ray protons interact with gas at the mean density of
the ISM in starbursts, then they lose energy rapidly
via inelastic proton-proton collisions with ambient nuclei.
This interaction generates both charged and neutral pions, which
subsequently produce secondary electrons and positrons, high energy
neutrinos, and photons, respectively.
For protons of GeV energy, the total inelastic proton-proton cross
section is $\approx$30\,mb (with inelasticity $\sim$0.5; e.g., Gaisser
1990)
so that the cosmic ray proton energy-loss timescale is
\beq
\tau_{pp}\approx 7\times10^7\,n^{-1}\,\,\,{\rm yr},
\label{taupp}
\eeq
where $n$ is the number density in units of cm$^{-3}$.  The
energy-loss timescale decreases by less than a factor of two
over six orders of magnitude in proton energy.

If the pion production timescale within a galaxy is shorter than the
escape timescale of relativistic protons, then the protons will lose
most of their energy to secondary photons, electron/positron pairs,
and neutrinos before escaping the galaxy.  The galaxy would then be a
cosmic ray proton calorimeter.

The escape timescale for cosmic ray protons is uncertain.  Both
energy-dependent diffusion and advection in a galactic outflow likely
contribute.  In the Galaxy, the diffusive escape timescale for 10 GeV
protons is inferred to be $\tau_{\rm diff}\sim3\times10^7$ yr near the
solar circle (Garcia-Munoz et al.~1977; Connell 1998).  Such protons
interact with a mean total column density of $\sim$$10$ g cm$^{-2}$
before escaping, and the column decreases at higher energies roughly
as $E^{-1/2}$ (e.g., Engelmann et al.~1990; Webber at al.~2003).
These observations imply that 10 GeV cosmic rays interact with matter
of density $n\sim0.2$ cm$^{-3}$, somewhat below the average density of
the ISM in the gas disk of the Milky Way.  Using equation
(\ref{taupp}) we infer that 10 GeV protons lose roughly $10 \%$ of
their energy to pion production before escaping.  Assuming that
diffusive losses proceed similarly in other star-forming galaxies,
these results imply that there is a critical gas surface density
$\Sigma_{\rm crit}$ above which protons will lose most of their energy
to pion production.  Taking the Milky Way's surface density to be
$\Sigma_{\rm MW_\odot} \approx2.5\times10^{-3}$ g cm$^{-2}$ at the
solar circle (e.g., Boulares \& Cox 1990) we estimate that \beq
\Sigma_{\rm crit}^{\rm diff} \sim 10 \, \Sigma_{\rm MW_\odot} \approx
0.03 \,\,{\rm g \,\,cm^{-2}}.
\label{sigmacrit_d}
\eeq

For starburst galaxies, the assumption that diffusion dominates the
escape of cosmic rays likely overestimates the escape timescale.  In
these systems escape is probably dominated by advection out of the gas
mid-plane in a starburst-driven wind (as, e.g., in M82; see Klein et
al.~1988, Seaquist \& Odegard 1991).  An estimate for the escape
timescale is then \beq \tau_{\rm wind}\approx h/V \approx 3\times10^5
h_{100}V_{300}^{-1}\,\,{\rm yr},
\label{tauwind}
\eeq where $h_{100}=h/100$ pc is the gas scale height and
$V_{300}=V/300$ km s$^{-1}$ is a typical wind velocity
(e.g., Martin 1999).
The condition $\tau_{pp}\lesssim\tau_{\rm wind}$ is then the criterion
for cosmic ray protons to lose a significant fraction of their energy
to pion production.  This criterion may also be written in terms of a
critical gas surface density;
\beq
\Sigma_{\rm crit}^{\rm wind}\approx0.3\,V_{300}\,\,{\rm g\,\,cm^{-2}}.
\label{sigmacrit}
\eeq

Although uncertain, equations (\ref{sigmacrit_d}) and
(\ref{sigmacrit}) imply that galaxies with $\Sigma_g \gtrsim 0.03-0.3$
g cm$^{-2}$ are likely to be proton calorimeters.  A number of local
starburst galaxies satisfy this criterion (see Table 1).

\section{$\gamma$-ray Emission from Starburst Galaxies \& \\ the
FIR-$\gamma$-ray Correlation}
\label{section:correlations}

\begin{figure*}
\centerline{\hbox{\psfig{file=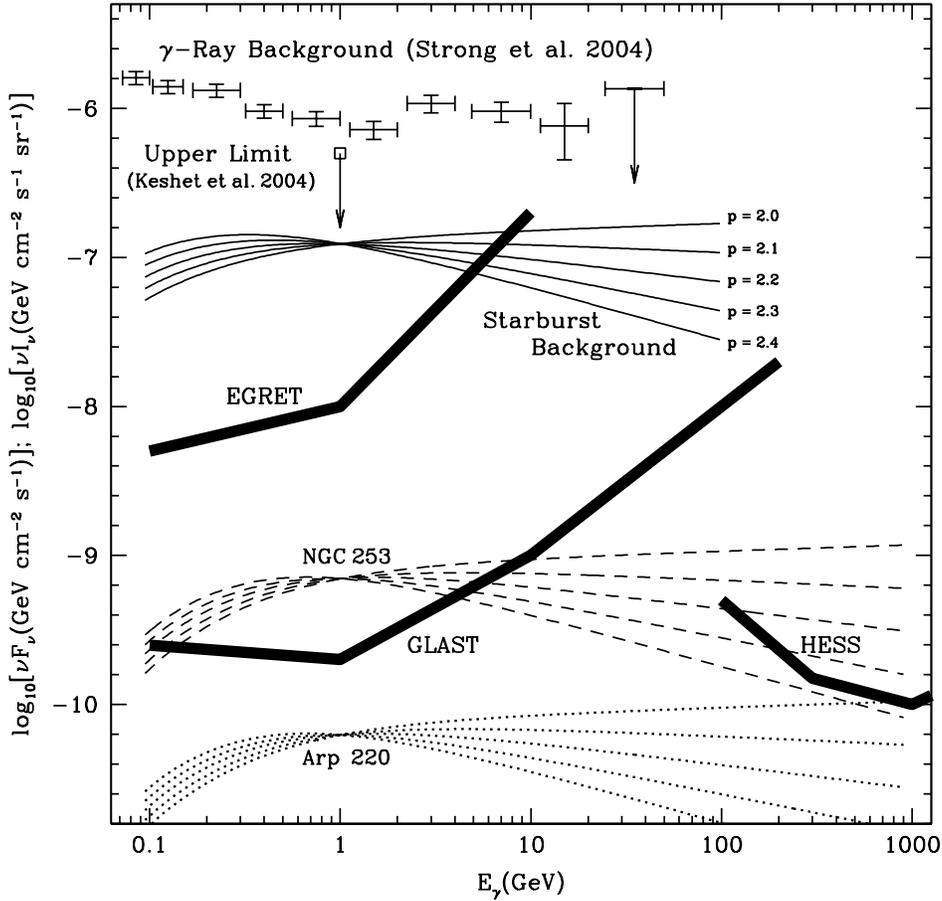,width=13cm}}}
\figcaption[x]{$\gamma$-ray flux and specific intensity as a function
of $\gamma$-ray energy. The data points with error bars show the
extra-galactic
$\gamma$-ray background as inferred by Strong et al.~(2004) from the
{\it EGRET} data.  The open square shows the upper limit
on the background derived by Keshet et al.~(2004) at 1 GeV.
The thick solid lines are the point source
sensitivities of {\it EGRET} and {\it GLAST} for a one year all-sky
survey and the {\it HESS} sensitivity curve for a 5$\sigma$ point
source detection using a 50 hour integration.  The solid lines show
the total starburst background calculated with $p=2.0-2.4$ and
assuming $\Omega_m=0.3$, $\Omega_\Lambda=0.7$, $H_0=71$ km s$^{-1}$
Mpc$^{-1}$,
$f(z)=\min[1,0.9\,z+0.1]$, and $\dot{\rho}_\star(z)$ from RSF2
in Porciani \& Madau (2001).   For this model $f_{\rm cal}\approx0.8$ (see
eq.~\ref{background}).
The expected signal from the starbursts NGC 253 ({\it dashed lines}) and
Arp 220 ({\it dotted lines}) with $p=2.0-2.4$ (see
\S\ref{section:calorimeter},
eqs.~[\ref{gamma_fir}] \& [\ref{fir_radio}]) are also shown. For other
starburst galaxies potentially visible by {\it GLAST} and {\it HESS},
see Table \ref{table:starburst}.  For both the $\gamma$-ray background
and the individual starbursts, the $p=2.1$ curves are normalized such that
the total integrated $\gamma$-ray luminosity is
$1.5\times10^{-4}\eta_{0.05}$
times the total integrated stellar luminosity (eq.~[\ref{gamma_fir}]).
Because of the constraints on the radio emission from starbursts,
all $p\ne2.1$ curves are normalized to the $p=2.1$ calculation at 1 GeV,
which implies that in our models, $p$ and $\eta$ are not independent
(see discussion in \S\ref{section:calorimeter}).  Note that the effects of
attenuation of the background spectrum above $\sim$10\,GeV have
been neglected (see, e.g., Madau \& Phinney 1996).
\label{fig:background}}
\end{figure*}

To estimate the $\gamma$-ray emission associated with pion production
in starbursts, we assume that the star formation rate ($\dot{M}_\star$) is
related to the total
IR luminosity ($L_{\rm TIR}$ $[8-1000]\mu$m; see, e.g., Calzetti et
al.~2000, Dale et al.~2001)
by $L_{\rm TIR}=\epsilon\dot{M}_\star c^2$ ($\epsilon$ is an IMF-dependent
constant), that the supernova rate per unit star formation, $\Gamma_{\rm
SN}$,
is a constant fraction of $\dot{M}_\star$, and that $5\eta_{0.05}\%$ of
the energy of each
supernova ($E_{51}=E_{\rm SN}/10^{51}$ ergs) is supplied to relativistic
protons.
With these assumptions, the total energy per unit time injected in cosmic
ray protons is
\beq
L_{{\rm CR}_p}\approx4.6\times10^{-4}L_{\rm
TIR}\,E_{51}\,\eta_{0.05}\,\beta_{17},
\label{lcr}
\eeq where $\beta_{17}=(\Gamma_{\rm SN}/\epsilon)/17$\,M$_\odot^{-1}$;
for continuous star formation over $\sim$10$^8$ yrs, this ratio has
only a very weak dependence on the assumed IMF, since high mass stars
dominate both the total luminosity and the supernova rate. In
addition, for star-formation timescales of $3 \times 10^7-10^9$\,yrs,
$\beta$ increases by only a factor of approximately 1.5, from
$\approx$15 to $\approx$23 (e.g., Leitherer et al.~1999).  We further
assume that starburst galaxies are proton calorimeters
($\tau_{pp}<\tau_{\rm esc}$, $\Sigma_g>\Sigma_{\rm crit}$) and --- for
the purposes of an analytic estimate --- that the $\gamma$-ray
spectrum is flat at and above GeV energies; this requires that cosmic
ray protons are injected with a spectral index $p\approx2$ (e.g.,
Blandford \& Eichler 1987).  Numerical calculations for different $p$
are shown in Figure \ref{fig:background} and discussed below.

Because the cross-sections for neutral- and charged-pion production
are similar, $\approx1/3$ of the proton energy goes into neutral
pions, which subsequently decay into $\gamma$-rays.  Thus, under the
assumption of proton calorimetry, the expected $\gamma$-ray luminosity
can be directly related to $L_{{\rm CR}_p}$ and $L_{\rm TIR}$
(eq.~[\ref{lcr}]). For a flat spectrum and for energies $\gtrsim$GeV,
we find that \beq \nu L_\nu ({\rm GeV}) \approx 1.5 \times 10^{-4}
\frac{\eta_{0.05}L_{\rm TIR}}{\ln(\gamma_{max})} \approx 10^{-5} \,
\eta_{0.05} \, L_{\rm TIR},
\label{gamma_fir}
\eeq where $\gamma_{max}(=10^6)$ is the assumed maximum Lorentz factor
of the accelerated protons and we have dropped the dependence here and
below on the ratio $\beta=(\Gamma_{\rm SN}/\epsilon)$.

Approximately $2/3$ of the energy in cosmic ray protons goes to
charged pions, which subsequently decay into high energy neutrinos and
electron/positron pairs. The pairs receive $\approx 1/4$ of the pion
energy (Schlickeiser 2002).  Thus $\approx 0.8 \, \eta_{0.05} \%$ of
the energy of each supernova goes into secondary electrons and
positrons.  Assuming that the synchrotron cooling time of the
secondary pairs is less than their escape time (V\"olk 1989; T06), we
find a FIR-radio correlation of the form \beq
\hspace*{-.1cm}\nu L_\nu ({\rm GHz}) \approx
7.5 \times 10^{-5}\frac{ \eta_{0.05} L_{\rm TIR}}{2\ln(\gamma_{max})}
\approx 2.5 \times 10^{-6} \, \eta_{0.05} \, L_{\rm TIR}.
\label{fir_radio}
\eeq The observed correlation is $\nu L_\nu(1.4\,\,{\rm GHz}) \approx
1.1\times10^{-6} L_{\rm TIR}$ (Yun et al.~2001).\footnote{$L_{\rm FIR}$,
as defined in Yun et al.~(2001) from the 60$\mu$m and
100$\mu$m IRAS fluxes (Helou et al.~1988), is generally a factor of
$\approx1.75$
smaller than $L_{\rm TIR}$ for starburst galaxies (see Calzetti et
al.~2000).}
For $\eta=0.05$, equation (\ref{fir_radio}) somewhat
over-predicts the observed FIR-radio correlation in starbursts.
However, this prediction neglects ionization, inverse Compton, and
bremsstrahlung losses, which may be competetive with synchrotron
losses in starbursts and will reduce the magnitude of
$\nu L_\nu ({\rm GHz}) /L_{\rm TIR}$.\footnote{Equation (\ref{fir_radio})
also neglects
the contribution to the radio emission from primary electrons
accelerated directly in supernova shocks.  Typical estimates for the
ratio of the total energy injected into protons and primary
electrons are $\sim$$20-40$ (e.g., Beck \& Krause 2005), in which
case secondary electrons/positrons likely dominate the radio
emission in starbursts (see T06; Rengarajan 2005).}
For example, in the models of T06, which simultaneously explain the
linearity of the
FIR-radio correlation and the radio spectra of starbursts (their
Figs.~3\,\&\,4), bremsstrahlung and ionization losses account for
$\approx$50\% of the energy lost by cosmic ray electrons.
Note that even if inverse Compton and bremsstrahlung losses completely
dominate ionization and synchrotron losses for secondary
electrons/positrons,
their contribution to the total X-ray and $\gamma$-ray emission from
starbursts
is at most 1/2 of the contribution from neutral-pion decay;  inverse
Compton
and bremsstrahlung will, however, dominate the emission below $\sim$100
MeV.

Equations (\ref{gamma_fir}) and (\ref{fir_radio}) highlight the
importance of the FIR-radio correlation for constraining the
$\gamma$-ray emission from starbursts.  Statistically, the
$\gamma$-ray emission from galaxies cannot exceed that predicted by
equation (\ref{gamma_fir}) without violating the radio constraints
implied by the FIR-radio correlation (eq.~\ref{fir_radio}).  The only way
out of this
conclusion is if ionization, inverse Compton, and bremsstrahlung
losses significantly exceed synchrotron losses for cosmic ray
electrons and positrons.  Although possible, it would be remarkable if
synchrotron losses were highly sub-dominant in starbursts and yet
somehow the linearity of the FIR-radio correlation is maintained, both
within different starburst galaxies and between starbursts and normal
star-forming galaxies.

%%%%%%%%%%%%%%%%%%%%%%%%%%%%%%%%%%%%%%%%%%%%%%%%%%%%%%%%%%%%%%%%%%%%%%%%%%%%
%%%%%%%%%%%%%%%%%%%%%%%%%%%%%%%%%%%%%%%%%%%%%%%%%%%%%%%%%%%%%%%%%%%%%%%%%%%%%

Equation (\ref{fir_radio}) is only consistent with the observed
FIR-radio correlation for $\eta \sim 0.05$, i.e., if $\sim$5\% of the
supernova energy is supplied to relativistic protons.  Although this
efficiency of cosmic ray production is reasonably consistent with the
energetics of galactic cosmic rays (e.g., Schlickeiser 2002) and with
the $\gamma$-ray luminosity of the Milky Way (Hunter et al.~1987),
some models of non-linear diffusive shock acceleration imply $\eta
\sim 0.5$ (e.g., Ellison \& Eichler 1984; Ellison et
al.~2004).\footnote{However, V\"olk et al.~(2003) argue that standard
spherical models of non-linear shock acceleration over-estimate the
efficiency of cosmic ray production by a factor of $\sim$5.}  If
indeed $\eta \gg 0.05$, then the FIR-radio correlation implies
that either starburst galaxies are not proton calorimeters or that
synchrotron losses are highly sub-dominant with respect to ionization,
inverse Compton, and bremsstrahlung losses.

Equations (\ref{gamma_fir}) and (\ref{fir_radio}) imply that the
$\gamma$-ray fluxes of the densest starbursts should be linearly
proportional to their radio and IR fluxes in the proton calorimeter
limit.  Table \ref{table:starburst} provides a sample of galaxies
chosen to have large IR fluxes and gas densities (see also Paglione et
al.\,1996; Blom et al.\,1999; Torres et al.~2004; Torres 2004;
Domingo-Santamar{\'{\i}}a\,\&\,Torres 2005; Torres \&
Domingo-Santamar{\'{\i}}a 2005).  For the latter, we have applied a
cut $\Sigma_g\ge0.02$\,g\,cm$^{-2}$, plausibly in the proton
calorimeter limit; the implied mean gas densities range from
$\sim$$20$\,cm$^{-3}$ for IC 342 to $\sim$$10^4$\,cm$^{-3}$ for Arp
220.  The systems in Table \ref{table:starburst} are listed in order
of decreasing TIR flux density, which corresponds to decreasing
$\gamma$-ray flux in the proton calorimeter limit
(eq. [\ref{gamma_fir}]).  Several systems just miss our cut on
$\Sigma_g$, including the nucleus of NGC 5236, NGC 4631, and NGC
3521.\footnote{ The Seyfert 2 galaxy NGC 1068 with
$\Sigma_g\approx0.1$ g cm$^{-2}$ (Helfer et al.~2003) would be fourth
in Table \ref{table:starburst} with $\nu F_\nu({\rm GeV}) \approx11.8
f \times10^{-11}$ GeV cm$^{-2}$ s$^{-1}$ where $f$ is the fraction of
the infrared emission due to the nuclear starburst rather than the
AGN.}  Because typical errors in $\Sigma_g$ are of order $\pm0.2$\,dex
(e.g., Kennicutt 1998) and because of uncertainty in $\Sigma_{\rm
crit}$ (\S\ref{section:calorimeter}), these systems may also prove to
be bright sources in the $\gamma$-ray sky.

The $\gamma$-ray fluxes predicted by equation (\ref{gamma_fir}) at 1
GeV from $L_{\rm TIR}$ are also provided in Table \ref{table:starburst}.
For comparison,
the expected sensitivity of {\it GLAST} at 1 GeV for a 1 year
integration is $\approx 2\times10^{-10}$ GeV cm$^{-2}$ s$^{-1}$ (see
Fig.~\ref{fig:background}).\footnote{See
http://people.roma2.infn.it/$\sim$glast/ for a comparison of
$\gamma$-ray telescope sensitivities.} Table
\ref{table:starburst} shows that there are 3 nearby starbursts above
this limit (M82, NGC 253,\& IC 342) and approximately 10 other
systems with predicted fluxes within a factor of 4 of {\it GLAST's}
detection limit.  We note that NGC 3690 and Arp 220 would be
particularly important for testing our predicted FIR-$\gamma$-ray
correlation over a large dynamic range in $L_{\rm TIR}$.

Figure \ref{fig:background} shows the $\gamma$-ray spectrum, as
calculated using the formalism of Aharonian \& Atoyan (2000) (their
\S4.1), as a function of energy for the specific cases of NGC 253
({\it dashed lines}) and Arp 220 ({\it dotted lines}), for
$p=2.0-2.4$, where $p$ is the slope of the proton injection spectrum.
We first specify $p$, and then we calculate the spectrum from equation
(22) of Aharonian \& Atoyan (2000), adopting their fit for the cross
section and their value of $K_\pi\approx0.17$ (see also the Appendix
of Domingo-Santamar{\'{\i}}a\,\&\,Torres 2005). We normalize the
resulting spectrum, such that the integral over all $\gamma$-ray
energies is set by the second equality in equation (\ref{gamma_fir}).
Comparing our analytic estimate in equation (\ref{gamma_fir}) with
this numerical calculation, we find that a nearly flat $p=2.1$ proton
spectrum gives the best agreement at GeV energies.  

In order to be consistent with the FIR-radio correlation as given in
the second equality in equation (\ref{fir_radio}),
all calculations with $p\ne2.1$ are normalized to the $p=2.1$ spectrum
at 1 GeV.  Note that this constraint implies that $p$ and $\eta$ are
{\it not independent}.  A constant flux at 1 GeV requires
$(\eta,p)\approx(0.09,2.0), \, (0.05,2.1), \, (0.03,2.2), \,
(0.025,2.3)$, and $(0.02,2.4)$.  Because the average magnetic field
strengths in starbursts may range from $B\sim0.1$\,mG (e.g., NGC 253
or M82) to $B\sim10$\,mG (e.g., Arp 220) (see T06 for a discussion),
and because the energy of GHz-emitting electrons/positrons is
proportional to $B^{-1/2}$, our normalization of the $p\ne2.1$ curves
at 1 GeV (rather than the proper normalization of the 1 GHz radio
emission) introduces a factor of $\approx 2$ uncertainty in the
predicted $\gamma$-ray emission.

As Figure \ref{fig:background} emphasizes, NGC 253 (and M82, see
Table \ref{table:starburst}) should be readily detected with {\it GLAST}.
Existing observations by {\it HESS} at TeV energies (Aharonian et
al.~2005) yield an upper limit of $<1.9\times10^{-12}$ ph cm$^{-2}$
s$^{-1}$ sr$^{-1}$ (for a point source; $E_\gamma>300$ GeV) for NGC
253 that is very close to our prediction for $p=2$, suggesting that
longer integrations with {\it HESS} should be able to detect NGC 253
or significantly constrain the high energy spectral slope
(see also Domingo-Santamar{\'{\i}}a \& Torres 2005; Torres \&
Domingo-Santamar{\'{\i}}a 2005).

In two cases --- NGC 253 and Arp 220 --- we can compare our estimate
for the $\gamma$-ray luminosity with more detailed calculations by
Torres (2004) and Domingo-Santamar{\'{\i}}a\,\&\,Torres (2005).  To
make an apposite comparison, it is necessary to scale our results to
their assumed paramaters.  Using $\eta=0.165$ and $p=2.2$, Torres
(2004) finds for Arp 220 that $\nu F_\nu({\rm
GeV})\approx3\times10^{-10}$\,\,GeV\,\,cm$^{-2}$\,\,s$^{-1}$.
Rescaling our $\eta\approx0.03$ and $p=2.2$ result, and adjusting for
our somewhat ($\approx20$\%) higher TIR flux, we find that $\nu
F_\nu({\rm GeV})\approx2.8\times10^{-10}$\,\,GeV\,\,cm$^{-2}$\,\,s$^{-1}$.  
For NGC 253, Domingo-Santamar{\'{\i}}a\,\&\,Torres (2005) estimate $\nu
F_\nu({\rm GeV})\approx2\times10^{-9}$\,\,GeV\,\,cm$^{-2}$\,\,s$^{-1}$
with $\eta=0.1$ and $p=2.2$.  Again scaling our results, we find that
$\nu F_\nu({\rm GeV})\approx2.3\times10^{-9}$\,\,GeV\,\,cm$^{-2}$\,\,s$^{-1}$ 
for their assumed parameters.  The correspondence between our predicted
fluxes and these more detailed calculations is thus excellent. These
studies also find, in agreement with our assumptions, that $\pi^0$
decay dominates $\gamma$-ray emission at GeV energies, that secondary
electrons and positrons dominate the radio emission at GHz
frequencies, and that both systems are safely in the proton and
electron/positron calorimeter limits; the cooling timescale is less
than the escape timescale.

The predicted fluxes for Arp 220 and NGC 253 above -- for Torres
(2004) and Domingo-Santamar{\'{\i}}a\,\&\,Torres's (2005) parameters
-- are a factor of $\approx 5$ and $3$ times higher than those given in
Table \ref{table:starburst} (primarily because of the larger values of
$\eta$). Given the constraints imposed by the FIR-radio correlation,
it is therefore perhaps surprising that Torres (2004) and
Domingo-Santamar{\'{\i}}a\,\&\,Torres's (2005) are able to fit the
radio emission from Arp 220 and NGC 253 with the significantly larger
values of $\eta$ they assumed.  As per the discussion after equation
(\ref{fir_radio}), this is only possible in the proton calorimeter
limit if ionization, bremsstrahlung, and/or inverse Compton losses
dominate synchrotron losses.  Indeed, for the individual nuclei of Arp
220 and for the inner starburst of NGC 253, Torres (2004) and
Domingo-Santamar{\'{\i}}a\,\&\,Torres's (2005) parameters (gas
densities, magnetic field strengths, etc.) imply that the ratio of the
synchrotron cooling timescale to the total electron energy loss
timescale (including all processes; dominated by ionization losses) is
$\sim$5.  Thus, their results are fully consistent with the arguments
of our paper. However, we believe that their predicted $\gamma$-ray
fluxes are optimistic because, as noted above, the tightness
and linearity of the FIR-radio correlation are difficult to explain if
synchrotron losses are highly sub-dominant in all star-forming
galaxies.

\section{The Starburst $\gamma$-Ray Background}
\label{section:background}

The total diffuse extra-galactic TIR background from star formation is
typically estimated to be $F^{\rm TIR}_{20}=F^{\rm
TIR}/20$\,nW\,\,m$^{-2}$\,\,sr$^{-1}$$\approx1$, with roughly half of
the contribution to the background coming from $z>1$ (e.g., Nagamine
et al.~2006).  Equation (\ref{gamma_fir}) for the $\gamma$-ray flux from
a given starburst in the proton calorimeter limit then implies a total
integrated $\gamma$-ray background of 
\beq
F_\gamma\approx1.4\times10^{-6}\eta_{0.05}f_{0.75}F^{\rm TIR}_{20}
\,\,\,\,{\rm GeV\,\,s^{-1}\,\,cm^{-2}\,\,sr^{-1}}
\label{background_total}
\eeq 
where $f_{0.75}=f_{\rm cal}/0.75$ and 
\beq 
f_{\rm cal}=\int_0^\infty \frac{\dot{\rho}_\star(z) f(z)\,dz}{[(1+z)^2E(z)]}/
\left\{\int_0^\infty
\frac{\dot{\rho}_\star(z)\,dz}{[(1+z)^2E(z)]}\right\} 
\eeq 
is the fraction of the TIR extragalactic background produced by starburst
galaxies that are in the proton calorimeter limit (i.e., $\Sigma_g >
\Sigma_{\rm crit}$; \S\ref{section:calorimeter}), $f(z)$ is that
fraction at every $z$, $\dot{\rho}_\star$ is the comoving star
formation rate density, and
$E(z)=[\Omega_m(1+z)^3+\Omega_\Lambda]^{1/2}$.  For a flat
$\gamma$-ray spectrum, equation (\ref{background_total}) implies a
specific intensity at GeV energies of
\beq 
\nu I_\nu ({\rm
GeV})\approx 10^{-7}\eta_{0.05}f_{0.75}F^{\rm TIR}_{20} \,\,\,\,{\rm
GeV\,\,s^{-1}\,\,cm^{-2}\,\,sr^{-1}}.
\label{background}
\eeq

The estimated $\gamma$-ray background depends sensitively on the fraction
of star formation that occurs in the proton calorimeter limit, i.e.,
on $f(z)$ and $f_{\rm cal}$.  We estimate $f(z=0) \sim 0.1$ from the
fraction of the local FIR and radio luminosity density produced by
starbursts (Yun et al.~2001).  Confirmation that local starbursts like
M82 and NGC 253 are $\gamma$-ray emitters at the level predicted in
Table \ref{table:starburst} would significantly solidify the
background predictions of equations (\ref{background_total}) and
(\ref{background}) by confirming that such starbursts are indeed
proton calorimeters.  At high redshift, a much larger fraction of star
formation occurs in high surface density systems likely to be
calorimeters.  For example, the strong evolution of the IR luminosity
function with redshift indicates that $F^{\rm TIR}$ is dominated by
luminous infrared galaxies at $z\sim1$ (Dole et al. 2006); these
systems are typical of those we expect to be proton calorimeters, in
which case $f(z) \sim 1$ at $z \sim 1$.  In addition, the typical gas
surface density of UV selected galaxies, which account for a
substantial fraction of the star formation at high redshift, is
roughly 
$\Sigma_g\sim0.06\,M_{10.5}\,R_{6\rm\,kpc}^{-2}$\,\,g\,\,cm$^{-2}$,
where $M_{10.5}$ is the inferred average gas mass in units of
$10^{10.5}$ M$_\odot$ and $R_{6\rm\,kpc}=R/6$ kpc is the half-light
radius (Erb et al.~2006ab).  This surface density is comparable to
$\Sigma_{\rm crit}$ estimated in \S\ref{section:calorimeter}.  Thus,
if we take $f(z\gtrsim1) \sim 1$ and $\dot{\rho}_\star(z)$ as in,
e.g., Porciani \& Madau (2001), we find that $f_{\rm cal} \sim 1$
(eq.~[\ref{background}]), with a significant contribution to the
$\gamma$-ray background coming from $z \approx 1-2$, so that \beq
F_\gamma\propto\nu I_\nu({\rm GeV})\propto f(z\approx1-2).
\label{f}
\eeq More specifically, taking $\Omega_m=0.3$ and
$\Omega_\Lambda=0.7$, and assuming a function $f(z)=\min(0.9z+0.1,1)$,
which smoothly interpolates from a small local starburst fraction
($f(z\approx0) \approx 0.1$) to an order unity starburst fraction at
high redshift, we find that $f_{\rm cal}\approx0.8$ for models RSF1,
RSF2, and RSF3 from Porciani \& Madau (2001) and for the ``Fossil''
model of Nagamine et al.~(2006).  These models also all have $F^{\rm
TIR}_{20} \approx 1$. As an alternative, assuming
$f(z)=\min[0.1(1+z)^3,0.8]$, in which the starburst fraction increases
roughly in proportion to $\dot{\rho}_\star(z)$, $f_{\rm
cal}\approx0.6$ for the same models.\footnote{Stecker (2006) presents
upper limits on the $\gamma$-ray and neutrino backgrounds from
starbursts that are significantly lower than our estimates here (see
\S\ref{section:discussion} for our neutrino background estimate).
There are two reasons for this difference.  First, Stecker advocates
that 22\% of the extragalactic TIR background is produced at $z >
1.2$ (see his Table 1) whereas we find that all models for
$\dot{\rho}_\star(z)$ that are consistent with observations of star
formation at high-$z$ and the local stellar mass density imply that
roughly half of the TIR background is produced at $z>1$ (e.g., RSF1,
RSF2, and RSF3 from Porciani \& Madau (2001), the ``Fossil'' model
from Nagamine et al.~2006, the dust-corrected star formation history
given by Schiminovich et al.~2005, and the analytic model given by
Hernquist \& Springel 2003).  Second, we argue that a fraction of
order unity of all star formation at $z\approx1$ occurs in starbursts
likely to be calorimeters (our $f(z)$; eq.~\ref{f}), whereas Stecker
quotes 13\% in the range $0.2\le z\le1.2$ and 60\% for $z>1.2$ (his
Table 1).  His step function model for $f(z)$ significantly
underestimates the contribution of $z \sim 1$ galaxies to the
$\gamma$-ray background.}

Using a different methodology, Pavlidou \& Fields (2002) calculate the
contribution to the $\gamma$-ray background from star-forming galaxies
and find a value for $\nu I_\nu$ at GeV energies that is $4-5$ times
higher than that given in equation (\ref{background}).  Pavlidou \&
Fields (2002) did not account for the constraints imposed by the
FIR-radio correlation, 
nor do they explicitly discuss the importance of proton calorimetry
and the essential competition between escape and pion losses
(\S\ref{section:calorimeter}).  By analogy with the Galaxy, Pavlidou
\& Fields (2002) assume that few GeV protons lose roughly 10\% of
their energy before escaping in all star-forming galaxies in the local
universe ($z\approx0$).  Because roughly 10\% of the local TIR
luminosity density is generated by starburst galaxies, and because we
assume that all starbursts are calorimeters, local starburst and
normal star-forming galaxies contribute roughly equally to the
$\gamma$-ray background at $z\approx0$.  As we have argued above,
however, starbursts are the dominant mode of star formation at high
$z$ and thus $z \sim 1-2$ starbursts dominate the contribution of
star-forming galaxies to the $\gamma$-ray background.

\section{Discussion}
\label{section:discussion}

Figure \ref{fig:background} shows the numerically calculated
$\gamma$-ray background as a function of $\gamma$-ray energy ({\it
solid lines}), for $p=2.0-2.4$ (see caption for details).  The
inferred background from {\it EGRET} from Strong et al.~(2004), as
well as the conflicting upper limit at 1 GeV derived by Keshet et
al.~(2004) are also shown.

Our results indicate that $\gamma$-ray production from pion decay in
starburst galaxies contributes significantly to the observed
background above 100 MeV. The considerable uncertainties in the
$\gamma$-ray background determination, primarily because of foreground
subtraction (see Keshet et al.~2004), complicate a more direct
comparison.  An important point of this paper is that the absolute
starburst $\gamma$-ray flux --- both that from individual galaxies and
that of the background --- cannot exceed that predicted in Figure
\ref{fig:background} and Table 1 without over-producing the radio
emission from secondary electrons and positrons produced in charged
pion decay (see \S\ref{section:correlations}, eq.~[\ref{fir_radio}]).
This conclusion is independent of whether starbursts are in fact
proton calorimeters (as we have assumed), and can only be circumvented
if ionization, inverse Compton, and bremsstrahlung losses
significantly exceed synchrotron losses for cosmic ray electrons and
positrons in starbursts; such a determination would put strong
constraints on the origin of the FIR-radio correlation (T06;
\S\ref{section:correlations}).  If detected by {\it GLAST}, the low
energy $\gamma$-ray emission from systems like NGC 253 and M82
(\S\ref{section:calorimeter}, Fig.~\ref{fig:background}, Table
\ref{table:starburst}), may help constrain the contribution of inverse
Compton and bremsstrahlung to electron/positron losses.

The primary physical uncertainty in our estimate of the $\gamma$-ray
fluxes from starbursts lies in whether cosmic rays do in fact interact
with gas at approximately the mean density of the ISM.  In particular,
given that galactic winds efficiently remove mass and metals from
galaxies (e.g., Heckman et al.~1990), it is unclear whether the cosmic
rays actually interact with the bulk of the ISM, which is required for
pion production to be significant (\S\ref{section:calorimeter}).
Detection of $\gamma$-rays from starbursts at the level predicted in this
paper
would thus provide an important constraint on the physics of the ISM in
starbursts
and on the coupling between supernovae and the dense ISM, which contains
most
of the mass.  In addition, the ionization and bremsstrahlung
energy-loss timescales for electrons/positrons are similar to the
proton-proton pion production timescale.  Thus, constraints on pion
production in starbursts via $\gamma$-ray observations directly
constrain the importance of ionization and bremsstrahlung losses for
shaping the radio emission from starburst galaxies (T06). By contrast,
detection of (or upper limits on) the $\gamma$-ray emission from
starbursts well below our predictions would rule out the hypothesis
that starbursts are proton calorimeters, although they would, in our
opinion, still be electron calorimeters. The alternative
possibility suggested by equation (\ref{gamma_fir}), that $\eta \ll 0.05$,
is unlikely, given the energetics of Galactic cosmic ray production
(e.g., Strong et al.~2004).

We note that the total IR background from the particular model of the
star formation history of the universe used to produce the
$\gamma$-ray background in Figure \ref{fig:background} is $F^{\rm
TIR}\approx2\times10^{-5}$ ergs cm$^{-2}$ s$^{-1}$ sr$^{-1}$, or
$\approx20$ nW m$^{-2}$ sr$^{-1}$, with approximately 80\% ($f_{\rm
cal}\approx0.8$) coming from starburst galaxies.  A complete
calculation, accounting for the contribution from the old stellar
population (which contributes significantly to the background at $z
\approx 0$), finds a total background in starlight that is roughly twice
as large (e.g., Nagamine et al.~2006).  Equation (\ref{fir_radio})
and, indeed, the existence of the FIR-radio correlation, implies that
the expected radio background is $\nu I_\nu({\rm
radio})\approx7.5\times10^{-10}\eta_{0.05}/[2\ln(\gamma_{max})]\approx2.7\times10^{-11}$
ergs\,cm$^{-2}$\,s$^{-1}$\,sr$^{-1}$.  A factor of two reduction in
$\nu I_\nu({\rm radio})$ has been included in this estimate as per the
discussion after equation (\ref{fir_radio}).

In addition to producing secondary electron/positron pairs and
$\gamma$-rays, pion production is also a significant source of
high-energy neutrinos, with $\approx 5\%$ of the proton cosmic ray energy
going into neutrinos.  Two-thirds of this energy goes to muon-type
neutrinos.
Therefore, for starburst galaxies that are
proton calorimeters, we expect a FIR-neutrino correlation of the form
$\nu L_\nu(\mu-{\rm neutrinos})=(2/3)\nu L_\nu({\rm
all\,\,neutrinos})\approx
1.5\times10^{-4}\,\eta_{0.05}L_{\rm TIR}/\ln(\gamma_{max})
\approx10^{-5}\,\eta_{0.05}L_{\rm TIR}$, identical to the $\gamma$-ray
luminosity (eq.~[\ref{gamma_fir}]).
The corresponding cumulative extra-galactic $\mu$-neutrino background
at GeV energies is then the same as given in equation (\ref{background}).
Using a very similar estimate, Loeb \& Waxman (2006) have recently argued
that starbursts are likely to be an important contributor to the high-energy
neutrino background.  Because the ratio of the $\gamma$-ray to neutrino
luminosity from pion decay depends only on the micro-physics of pion
production, the $\gamma$-ray fluxes from starbursts predicted in this
paper will provide a crucial calibration of the expected flux of
extra-galactic high energy neutrinos.

\acknowledgments

T.A.T.~thanks Uri Keshet for a critical reading of the text and 
stimulating conversation.
E.Q.~is supported in part by NASA grant NNG06GI68G, an Alfred P.~Sloan
Fellowship, and the David and Lucile Packard Foundation.  E.W.~thanks
the Institute for Advanced Study and is supported in part by ISF and
AEC grants.  We thank the anonymous referee for a prompt and useful
report.

\begin{table}
\begin{scriptsize}
\begin{center}
\caption{Properties of Local Starburst \& Star-Forming Galaxies
\label{table:starburst}}

\begin{tabular}{lccccccccccc}
\hline \hline

\\

\multicolumn{1}{c}{Object Name} &
\multicolumn{1}{c}{$D$\tablenotemark{a}} &
\multicolumn{1}{c}{$\Sigma_g$\tablenotemark{b}} &
\multicolumn{1}{c}{$S_{60\mu\hspace{-.025cm}{\rm m}}$\tablenotemark{c}}&
\multicolumn{1}{c}{$S_{100\mu\hspace{-.025cm}{\rm m}}$\tablenotemark{d}}&
\multicolumn{1}{c}{$S_{{\rm TIR}}$\tablenotemark{e}}&
\multicolumn{1}{c}{$S_{1.4\,{\rm GHz}}$\tablenotemark{f}}&
\multicolumn{1}{c}{$\nu F_\nu({\rm GeV})$\tablenotemark{g}} &
\multicolumn{1}{c}{Data}\\

\multicolumn{1}{c}{} &
\multicolumn{1}{c}{(Mpc)} &
\multicolumn{1}{c}{$({\rm g\,cm^{-2}})$} &
\multicolumn{1}{c}{(Jy)} &
\multicolumn{1}{c}{(Jy)} &
\multicolumn{1}{c}{(Jy)} &
\multicolumn{1}{c}{(mJy)} &
\multicolumn{1}{c}{($10^{-11}$\,GeV\,cm$^{-2}$\,s$^{-1}$)} &
\multicolumn{1}{c}{Refs.} \\

\\
\hline
\\
NGC 3034 (M82)...          &  3.3 & 0.69  &  1313.5  & 1355.4 &  8302.4 &
7657.0 & 73  &  4,2,3 \\
NGC 253................    &  3.5 & 0.47  &   997.7  & 1857.8 &  7755.8 &
5594.0 & 68\tablenotemark{h}  &  1,2,3 \\
IC 342.................... &  4.4 & 0.02  &   255.9  &  660.7 &  2311.6 &
2250.0 & 20  &  1,2,3 \\
NGC 2146..............     & 12.6 & 0.14  &   144.5  &  204.2 &  1009.8 &
1074.6 & 8.9 &  1,2,3 \\
NGC 3690..............     & 42.2 & 0.04  &   119.7  &  118.6 &  748.0  &
678.3  & 6.6 &  4,2,3 \\
NGC 1808..............     & 14.2 & 0.09  &   104.5  &  147.2 &  729.4  &
529.0  & 6.4 &  1,2,3 \\
NGC 1365..............     & 18.6 & 0.09  &    84.1  &  185.4 &  704.2  &
530.0  & 6.2 &  7,8,3 \\
Arp 220..................  & 76.6 & 12.0  &   107.4  &  118.3 &  691.9  &
326.9  & 6.1\tablenotemark{i} &  4,2,3 \\
NGC 891................    &  7.4 & 0.09  &    61.1  &  198.6 &  623.4  &
701.0  & 5.5 &  6,2,3 \\
NGC 3627..............     & 10.2 & 0.04  &    56.2  &  144.9 &  507.3  &
458.0  & 4.5 &  4,2,3 \\
NGC 660................    & 12.0 & 0.08  &    65.5  &  103.8 &  477.4  &
387.0  & 4.2 &  4,2,3 \\
NGC 3628..............     & 11.2 & 0.02  &    48.4  &  121.9 &  431.9  &
525.0  & 3.8 &   10,3 \\
NGC 1097..............     & 18.0 & 0.10  &    46.7  &  116.1 &  414.0  &
415.0  & 3.6 &  4,2,3 \\
NGC 3079..............     & 15.9 & 3.7   &    50.2  &  103.5 &  407.8  &
849.0  & 3.6 &  4,2,3 \\
\\
\hline
\hline
\end{tabular}
\end{center}
\end{scriptsize}

\tablenotetext{a}{Distance. In most cases, distances computed with $H=71$
km s$^{-1}$ Mpc$^{-1}$.
In some cases, distances are taken from the literature.
$^{\rm b}$Gas surface density.  Typical uncertainty in this quantity is
$\pm0.2$\,dex (e.g., Kennicutt 1998).
$^{\rm c}$IRAS flux density at 60 $\mu$m.
$^{\rm d}$IRAS flux density at 100 $\mu$m.
$^{\rm e}$Total IR flux density defined by $\log_{10}(S_{\rm TIR}) \equiv
\log_{10}(2.58S_{60\mu\hspace{-.025cm}{\rm
m}}+S_{100\mu\hspace{-.025cm}{\rm
m}})+\log_{10}(1.75)$, where the factor of 1.75 relates the total
$[8-1000]$$\mu$m flux density to the $[40-120]$$\mu$m flux density
(Calzetti et al.~2000; Helou et al.~1988).  For any individual system
listed here,
this bolometric correction is uncertain at the factor of $\sim$1.5 level
(see  Bell 2003; Dale et al.~2001).
$^{\rm f}$Radio flux density at 1.4\,GHz.
$^{\rm g}$Predicted $\gamma$-ray flux at 1 GeV,
scaled from $S_{\rm TIR}$ according to eq.~(\ref{gamma_fir}),
assuming $\eta=0.05$ and $\gamma_{max}=10^6$.
$^{h}$See discussion of comparison with
Domingo-Santamar{\'{\i}}a\,\&\,Torres (2005) in
\S\ref{section:calorimeter}.
$^{i}$See discussion of comparison with Torres (2004) in
\S\ref{section:calorimeter}.
}

\tablerefs{(1) Condon (1987); (2) Kennicutt (1998); (3) Yun, Reddy, \&
Condon (2001); (4) Condon et al.~(1990); (5) Golla \& Wielebinski (1994);
(6) Condon et al.~(1996); (7) Sandqvist et al.~(1995); (8) Sandqvist
(1999); (10) Paglione et al.~(2001); (11) Condon et al.~(1998)}

\end{table}


\begin{thebibliography}{}

\bibitem[Aharonian \& Atoyan(2000)]{2000A&A...362..937A}
Aharonian, F.~A., \& Atoyan, A.~M.\ 2000, \aap, 362, 937

\bibitem[Aharonian et al.(2005)]{2005A&A...437L...7A}
Aharonian, F., et al.\ 2005, \aap, 437, L7

\bibitem[Beck \& Krause(2005)]{2005AN....326..414B}
Beck, R., \& Krause, M.\ 2005, Astronomical Notes, 326, 414

\bibitem[Bell(2003)]{2003ApJ...586..794B} Bell, E.~F.\ 2003, \apj, 586,
794

\bibitem[Boulares \& Cox(1990)]{1990ApJ...365..544B}
Boulares, A., \& Cox, D.~P.\ 1990, \apj, 365, 544

\bibitem[Blandford \& Eichler(1987)]{1987PhR...154....1B}
Blandford, R., \& Eichler, D.\ 1987, \physrep, 154, 1

\bibitem[Blom et al.(1999)]{1999ApJ...516..744B} Blom, J.~J., Paglione,
T.~A.~D., \& Carrami{\~n}ana, A.\ 1999, \apj, 516, 744

\bibitem[Calzetti et al.(2000)]{2000ApJ...533..682C}
Calzetti, D., Armus, L., Bohlin, R.~C., Kinney, A.~L., Koornneef, J., \&
Storchi-Bergmann, T.\
2000, \apj, 533, 682

\bibitem[Cillis et al.(2005)]{2005ApJ...621..139C}
Cillis, A.~N., Torres, D.~F., \& Reimer, O.\ 2005, \apj, 621, 139

\bibitem[Colafrancesco \& Blasi(1998)]{1998APh.....9..227C}
Colafrancesco, S., \& Blasi, P.\ 1998, Astroparticle Physics, 9, 227

\bibitem[Condon(1987)]{1987ApJS...65..485C}
Condon, J.~J.\ 1987, \apjs, 65, 485

\bibitem[Condon \& Broderick(1988)]{1988AJ.....96...30C}
Condon, J.~J., \& Broderick, J.~J.\ 1988, \aj, 96, 30

\bibitem[Condon et al.(1990)]{1990ApJS...73..359C}
Condon, J.~J., Helou, G., Sanders, D.~B., \& Soifer, B.~T.\ 1990, \apjs,
73, 359

\bibitem[Condon et al.(1991)]{1991ApJ...378...65C}
Condon, J.~J., Huang, Z.-P., Yin, Q.~F., \& Thuan, T.~X.\ 1991, \apj, 378,
65

\bibitem[Condon(1992)]{1992ARA&A..30..575C}
Condon, J.~J.\ 1992, \araa, 30, 575

\bibitem[Condon et al.(1996)]{1996ApJS..103...81C}
Condon, J.~J., Helou, G., Sanders, D.~B., \& Soifer, B.~T.\ 1996, \apjs,
103, 81

\bibitem[Condon et al.(1998)]{1998AJ....115.1693C}
Condon, J.~J., et al.\ 1998, \aj, 115, 1693

\bibitem[Connell(1998)]{1998ApJ...501L..59C}
Connell, J.~J.\ 1998, \apjl, 501, L59

\bibitem[Dar \& Shaviv(1995)]{1995PhRvL..75.3052D}
Dar, A., \& Shaviv, N.~J.\ 1995, Physical Review Letters, 75, 3052

\bibitem[Dole et al.(2006)]{2006A&A...451..417D} Dole, H., et al.\ 2006, 
\aap, 451, 417 

\bibitem[Domingo-Santamar{\'{\i}}a \& Torres(2005)]{2005A&A...444..403D}
Domingo-Santamar{\'{\i}}a, E., \& Torres, D.~F.\ 2005, \aap, 444, 403

\bibitem[Ellison et al.(2004)]{2004A&A...413..189E} Ellison, D.~C.,
Decourchelle, A., \& Ballet, J.\ 2004, \aap, 413, 189

\bibitem[Ellison \& Eichler(1984)]{1984ApJ...286..691E} Ellison, D.~C., \&
Eichler, D.\ 1984, \apj, 286, 691

\bibitem[Els{\"a}sser \& Mannheim(2005)]{2005PhRvL..94q1302E}
Els{\"a}sser,
D., \& Mannheim, K.\ 2005, Physical Review Letters, 94, 171302

\bibitem[Gaisser (1990)]{gaisser}
Gaisser, T.~K.~1990, Cosmic Rays and Particle Physics, (Cambridge:
Cambridge University Press)

\bibitem[Garcia-Munoz et al.(1977)]{1977ApJ...217..859G}
Garcia-Munoz, M., Mason, G.~M., \& Simpson, J.~A.\ 1977, \apj, 217, 859

\bibitem[Golla \& Wielebinski(1994)]{1994A&A...286..733G} Golla, G., \&
Wielebinski, R.\ 1994, \aap, 286, 733

\bibitem[Heckman et al.(1990)]{1990ApJS...74..833H} Heckman, T.~M., Armus,
L., \& Miley, G.~K.\ 1990, \apjs, 74, 833

\bibitem[Helfer et al.(2003)]{2003ApJS..145..259H}
Helfer, T.~T., et al.~2003, \apjs, 145, 259

\bibitem[Helou et al.(1985)]{1985ApJ...298L...7H}
Helou, G., Soifer, B.~T., \& Rowan-Robinson, M.\ 1985, \apjl, 298, L7

\bibitem[Helou et al.(1988)]{1988ApJS...68..151H} Helou, G., Khan, I.~R.,
Malek, L., \& Boehmer, L.\ 1988, \apjs, 68, 151

\bibitem[Hernquist \& Springel(2003)]{2003MNRAS.341.1253H} Hernquist, L., 
\& Springel, V.\ 2003, \mnras, 341, 1253 

\bibitem[Hunter et al.(1997)]{1997ApJ...481..205H} Hunter, S.~D., et al.\
1997, \apj, 481, 205

\bibitem[Kennicutt(1998)]{1998ApJ...498..541K}
Kennicutt, R.~C.\ 1998, \apj, 498, 541

\bibitem[Keshet et al.(2003)]{2003ApJ...585..128K} Keshet, U., Waxman, E., 
Loeb, A., Springel, V., \& Hernquist, L.\ 2003, \apj, 585, 128 

\bibitem[Keshet et al.(2004)]{2004JCAP...04..006K} Keshet, U., Waxman, E.,
\& Loeb, A.\ 2004, JCAP, 4, 6

\bibitem[Klein et al.(1988)]{1988A&A...190...41K}
Klein, U., Wielebinski, R., \& Morsi, H.~W.\ 1988, \aap, 190, 41

\bibitem[Leitherer et al.(1999)]{1999ApJS..123....3L}
Leitherer, C., et al.\ 1999, \apjs, 123, 3

\bibitem[Loeb \& Waxman(2000)]{2000Natur.405..156L} Loeb, A., \& Waxman,
E.\ 2000, \nat, 405, 156

\bibitem[Loeb \& Waxman(2006)]{2006JCAP...05..003L} Loeb, A., \& Waxman,
E.\ 2006, JCAP, 5, 3

\bibitem[Madau \& Phinney(1996)]{1996ApJ...456..124M}
Madau, P., \& Phinney, E.~S.\ 1996, \apj, 456, 124

\bibitem[Martin(2005)]{2005ApJ...621..227M}
Martin, C.~L.\ 2005, \apj, 621, 227

\bibitem[Miniati(2002)]{2002MNRAS.337..199M}
Miniati, F.\ 2002, \mnras, 337, 199

\bibitem[Murgia et al.(2005)]{2005A&A...437..389M} Murgia, M., Helfer, 
T.~T., Ekers, R., Blitz, L., Moscadelli, L., Wong, T., \& Paladino, R.\ 
2005, \aap, 437, 389

\bibitem[Murphy et al.(2006)]{2006ApJ...638..157M} Murphy, E.~J., et al.\ 
2006, \apj, 638, 157 

\bibitem[Nagamine et al.(2006)]{2006astro.ph..3257N} Nagamine, K.,
Ostriker, J.~P., Fukugita, M., \& Cen, R.\ 2006, arXiv:astro-ph/0603257

\bibitem[Niklas \& Beck(1997)]{1997A&A...320...54N} Niklas, S., \& Beck, 
R.\ 1997, \aap, 320, 54 

\bibitem[Paglione et al.(1996)]{1996ApJ...460..295P} Paglione, T.~A.~D.,
Marscher, A.~P., Jackson, J.~M., \& Bertsch, D.~L.\ 1996, \apj, 460, 295

\bibitem[Pavlidou \& Fields(2002)]{2002ApJ...575L...5P} Pavlidou, V., \&
Fields, B.~D.\ 2002, \apjl, 575, L5

\bibitem[Porciani \& Madau(2001)]{2001ApJ...548..522P}
Porciani, C., \& Madau, P.\ 2001, \apj, 548, 522

\bibitem[Schiminovich et al.(2005)]{2005ApJ...619L..47S} Schiminovich, D., 
et al.\ 2005, \apjl, 619, L47 

\bibitem[Seaquist \& Odegard(1991)]{1991ApJ...369..320S}
Seaquist, E.~R., \& Odegard, N.\ 1991, \apj, 369, 320

\bibitem[Sreekumar et al.(1998)]{1998ApJ...494..523S} Sreekumar, P., et
al.\ 1998, \apj, 494, 523

\bibitem[Stecker \& Salamon(1996)]{1996ApJ...464..600S} Stecker, F.~W., \&
Salamon, M.~H.\ 1996, \apj, 464, 600

\bibitem[Stecker (2006)]{stecker}
Stecker, F.~2006, arXiv:astro-ph/0607197 

\bibitem[Strong et al.(2000)]{2000ApJ...537..763S}
Strong, A.~W., Moskalenko, I.~V., \& Reimer, O.\ 2000, \apj, 537, 763

\bibitem[Strong et al.(2004)]{2004ApJ...613..956S} Strong, A.~W.,
Moskalenko, I.~V., \& Reimer, O.\ 2004, \apj, 613, 956

\bibitem[Thompson et al.(2006)]{2006ApJ...645..186T} Thompson, T.~A.,
Quataert, E., Waxman, E., Murray, N., \& Martin, C.~L.\ 2006, \apj, 645,
186

\bibitem[Torres et al.(2004)]{2004ApJ...607L..99T}
Torres, D.~F., Reimer, O., Domingo-Santamar{\'{\i}}a, E., \& Digel, S.~W.\
2004, \apjl, 607, L99

\bibitem[Torres (2004)]{torres_arp220}
Torres, D.~F., 2004, ApJ, 617, 966

\bibitem[Torres \& Domingo-Santamar{\'{\i}}a(2005)]{2005MPLA...20.2827T}
Torres, D.~F., \& Domingo-Santamar{\'{\i}}a, E.\ 2005, Modern Physics
Letters A, 20, 2827

\bibitem[Ullio et al.(2002)]{2002PhRvD..66l3502U} Ullio, P.,
Bergstr{\"o}m,
L., Edsj{\"o}, J., \& Lacey, C.\ 2002, \prd, 66, 123502

\bibitem[V\"olk(1989)]{1989A&A...218...67V}
V\"olk, H.~J.\ 1989, \aap, 218, 67

\bibitem[V{\"o}lk et al.(2003)]{2003A&A...409..563V} V{\"o}lk, H.~J.,
Berezhko, E.~G., \& Ksenofontov, L.~T.\ 2003, \aap, 409, 563

\bibitem[Yun et al.(2001)]{2001ApJ...554..803Y}
Yun, M.~S., Reddy, N.~A., \& Condon, J.~J.\ 2001, \apj, 554, 803

\end{thebibliography}
\end{document}